# Mechanism of cathodic protection of iron and steel in porous media

Federico Martinelli-Orlando[1], Shishir Mundra[1], Ueli M. Angst[1*]

[1]Institute for Building Materials (IfB), ETH Zurich, Laura-Hezner Weg 7, CH-8093 Zurich, Switzerland

*Corresponding author: uangst@ethz.ch

## Abstract

Cathodic protection (CP) was introduced two centuries ago and since has found widespread application in protecting structures such as pipelines, offshore installations, and bridges from corrosion. Despite its extensive use, the fundamental working mechanism of CP remains debated, particularly for metals in porous media such as soil. Here, we offer resolution to the long-standing debate by employing in-situ and ex-situ characterisation techniques coupled with electrochemical measurements to characterise the spatio-temporal changes occurring at the steel-electrolyte interface. We show that upon CP, the interfacial electrolyte undergoes alkalinisation and deoxygenation, and that depending on polarisation conditions, an iron oxide film can simultaneously form on the steel surface. We further demonstrate that these changes in interfacial electrolyte chemistry and steel surface state result in altered anodic and cathodic reactions and their kinetics. We propose a mechanism of CP that integrates the long debated theories, based on both concentration and activation polarisation, complimentarily. Implications of this coherent scientific understanding for enhancing corrosion protection technologies and the safe, economic, and environmental-friendly operation of critical steel-based infrastructures are discussed.

# Introduction

Exactly 200 years ago, Sir Humphry Davy, serving as President of the Royal Society at that time, implemented cathodic protection (CP) as a method to combat metal corrosion for first time on the basis of scientific principles[1–3]. Capitalising on the late 18th century findings of Galvani and Volta, Davy's experiments showed that coupling sacrificial anodes consisting of relatively unnoble metals with a more noble metal can protect the latter from corrosion. These findings rapidly led to the large-scale application of CP technology to protect corrosion of copper sheetings of ship hulls of the Royal Navy. While the corrosion of copper in seawater could be effectively mitigated, the protection current caused side effects that greatly impeded the practicality of utilising the technology on ships. Nowadays, it is well understood that these side effects are related to the changes in interfacial seawater chemistry, including an increase in the pH, resulting from the electrochemical reactions occuring on the protected metal surface, and ultimately enabling the formation of mineral deposits and thus promoting fouling[2,4]. In the early 19th century, Davy was not able to explain these side effects[2] and failed to resolve them, with serious consequences for the fleet of the Royal Navy and even the relationship between science and society as a whole[2,3]. Today, the formation of mineral scales upon cathodic protection is widely accepted, and a range of studies have shed light on the parameters controlling the deposition of mineral scales and their properties, especially for cathodically protected steel in seawater. These parameters include temperature, seawater chemistry (Mg/Ca ratio, pH, etc.), water flow rate, and the extent of cathodic polarisation[5–7].

In the early 20th century, CP was increasingly applied to protect steel structures in soil. By now, CP of ground-buried steel structures such as pipes and tank bottoms is among the most widely used corrosion protection technologies[8–13], and even required by law in various countries for certain safety-relevant infrastructure such as high-pressure gas lines. In soil, similar electrochemical reactions and changes in electrolyte chemistry as in seawater are expected to occur, although a major difference is that calcareous deposits in soil cannot always form, especially in soft water environments[11]. Despite the widespread adoption of CP in engineering practices and the long history of CP, fundamental questions about the underlying working mechanism are still under debate[10–12,14–18]. Divergent theories revolve around two main aspects: the sole consideration of corrosion kinetics[17,19] on the one hand, and the hypotheses surrounding the formation of iron oxide films on the steel due to the aforementioned changes in electrolyte chemistry[18,20–31], on the other hand. On both sides, scholars have mainly argued on the basis of theoretical reasoning and indirect measurements, while thorough investigations about changes in the surface state of the steel electrode (formation of oxide film or not) and the adjacent medium under CP are lacking.

This lack of fundamental and coherent understanding presents a challenge for devising scientifically sound engineering practices, namely technical guidelines and standards, which are central for the safe, economic, and environmental-friendly operation of various critical steel-based structures, such as in the energy and construction sector. This problem becomes a particularly pressing issue due to the enormous socio-economic importance of corrosion of infrastructure[32]. In the context of ageing infrastructure in industrialised countries[33,34], knowledge-based engineering practices are needed to ensure safety and to mitigate the ever-increasing costs of repairing and replacing deteriorating infrastructures. Moreover, CP can play an important role in reducing the environmental impact of infrastructure globally[8,35,36], notably by preventing losses of gas, oil, or water from pipelines and by extending the lifespan of structures.

While CP of steel in flowing electrolytes such as seawater is largely understood, CP of steel in porous media such as soil or concrete, is comparatively poorly understood, because of additional degrees of complexity stemming from aspects such as the tortuous nature of the system – limiting mass transport processes such as ion or oxygen diffusion and advection, or because of the variability in the electrolyte chemistry encountered in these porous media. For these reasons, this work aimed at unravelling the mechanism of CP of steel in a porous media, such as that of soil. Here, we use a comprehensive approach utilising in-situ and ex-situ spectroscopic and microscopic techniques, combined with electrochemical characterisation, to study the changes at the metal-electrolyte interface, and particularly at the metal surface, during cathodic polarisation.

# Results

*Time-resolved in-situ characterisation*

We potentiostatically polarised carbon steel specimens immersed in electrolyte to different cathodic or ON-potentials ($E_{ON}$) and recorded the cathodic current density (or protection current density, $i_{prot}$). This current density $i_{prot}$ was found to drastically increase with more negative $E_{ON}$ (Figure 1a). It is noticeable that in the simulated soil solution, $i_{prot}$ decreased over time, which indicates time-dependent changes in electrolyte chemistry or electrode reaction kinetics. Within approx. 10 h, however, $i_{prot}$ tended to stabilise, reaching steady-state values of ~ 4 A/m$^2$ (at -1.2 $V_{SSE}$), ~ 0.4 A/m$^2$ (at -1.0 $V_{SSE}$), ~ 0.25 A/m$^2$ (at -0.9 $V_{SSE}$), and ~ 0.1 A/m$^2$ (at -0.8 $V_{SSE}$).

The application of $i_{prot}$ enhances the cathodic reactions occurring at the steel surface, in particular the oxygen reduction reaction (ORR) and the hydrogen evolution reaction (HER). According to these reactions, the local pH at the metal-solution interface is expected to increase. We quantified this increase in pH in simulated soil solutions in a separate experimental setup. Small and minimally invasive pH sensors where installed at a given distance from the carbon steel electrode to monitor the time-dependent pH changes in simulated soil solution upon polarisation. Figure 1b shows that the pH in the electrolyte in vicinity of the metal surface increased. At a distance of 6 mm from the polarised steel, the pH of the simulated soil solutions ranged between pH 8-10 (at -0.92 $V_{SSE}$) and pH 10-13 (at -1.10 $V_{SSE}$) within a few hours of polarisation. The pH profiles shown in Figure 1c suggest even higher pH values at the steel surface. These observations of alkalinsation are qualitatively in agreement with literature[16,22–25,37,38]. To further corroborate the electrochemistry-driven alterations in near-surface electrolyte chemistry, the concentration of dissolved oxygen was measured with small sensors positioned at distance 6 mm from the steel surface in the same setup where pH measurements were done. Figure 1d shows that oxygen is rapidly consumed upon application of $i_{prot}$, supporting the aforementioned hypothesis that oxygen reduction reaction (ORR) occurs, and then, once the dissolved oxygen concentration has decreased to very low levels, the hydrogen evolution reaction (HER) may dominate.

For comparison with the initially pH near-neutral simulated soil solution, we exposed carbon steel directly (t = 0) to an alkaline solution with pH~13 and potentiostatically polarised to $E_{ON}$ = -0.8 $V_{SSE}$. In this case, $i_{prot}$ showed less pronounced changes over time, apart from a relatively small decrease within the first hour, and stabilised at a value of ~ 0.2 A/m$^2$ (Figure 1a). Since the electrolyte used had an elevated concentration of OH$^-$ (~0.1 M), the application of the cathodic current in the order of 0.25 A/m$^2$ is not expected to increase the pH significantly in the vicinity of the sample. It is interesting that $i_{prot}$ reached a comparable value in alkaline (pH~13) solution after ~1 h and in initially pH near-neutral simulated soil solution after ~10 h.

According to well-established theory[39], under alkaline conditions and with potentials more noble than the reversible potential of the Fe/Fe(II) electrode, the formation of an iron oxide film is expected. We tested this hypothesis of oxide film formation upon cathodic polarisation by measuring in-situ the absorption on the surface of carbon steel samples for a duration of 30 min, using UV-Vis spectroscopy. In simulated soil solution of initial near-neutral pH and polarised at $E_{ON}$ = -0.9 $V_{SSE}$, the absorption spectra (Figure 1e) increased over the experimental time, particularly in the UV region (wavelengths ($\lambda$) lower than 400 nm, also in line with observations in the iterature[34-36]) and with a characteristic peak at $\lambda$ = 380 nm. Similarly, in alkaline solution (pH~13) polarised at $E_{ON}$ = -0.8 $V_{SSE}$ (Figure 1e), the absorption on the steel surface showed a peak at $\lambda$ =380 nm. This suggests comparable surface states for steel cathodically polarised in an initially alkaline electrolyte, and cathodically polarised steel in an initially pH near-neutral electrolyte (that transformed to an alkaline electrolyte upon polarisation at $E_{ON}$ = -0.9 $V_{SSE}$).

Additionally, we monitored the film formed on the steel surface over time, by conducting in-situ photometric reflectance measurements (PRM) at $\lambda$ = 380 nm. At $E_{ON}$ = -0.9 $V_{SSE}$ in simulated soil solution, the absorption of the film was found to increase during the entire course of the experiment (Figure 1d), indicative of film growth. When the steel was exposed to an already initially highly alkaline electrolyte, the adsorption increased more rapidly than in the initially pH near-neutral simulated soil solution at $E_{ON}$ = -0.9 $V_{SSE}$, however, after 30 min comparable adsorption (suggesting comparable film thickness) was observed. In the case of polarisation at $E_{ON}$ = -1.2 $V_{SSE}$ in simulated soil solution, the formation of an oxide film was not evident over the entire course of the experiment. This may be expected from the Pourbaix diagram (Supplementary Figure 5) for iron that reveals that such negative polarisation lies within the immunity domain[46].

Given the similarity in the $i_{prot}$ and absorption data from UV-Vis spectroscopy and PRM for steel immersed in an already initially alkaline (pH~13) solution, polarised at $E_{ON}$ = -0.8 $V_{SSE}$, and steel immersed in simulated soil solution (initial pH 7.3), polarised at $E_{ON}$ = -0.9 $V_{SSE}$, we would be reasonable in expecting the iron oxide film formed on the samples exposed to the two electrolytes being similar in composition and thickness for the considered experimental time and conditions.

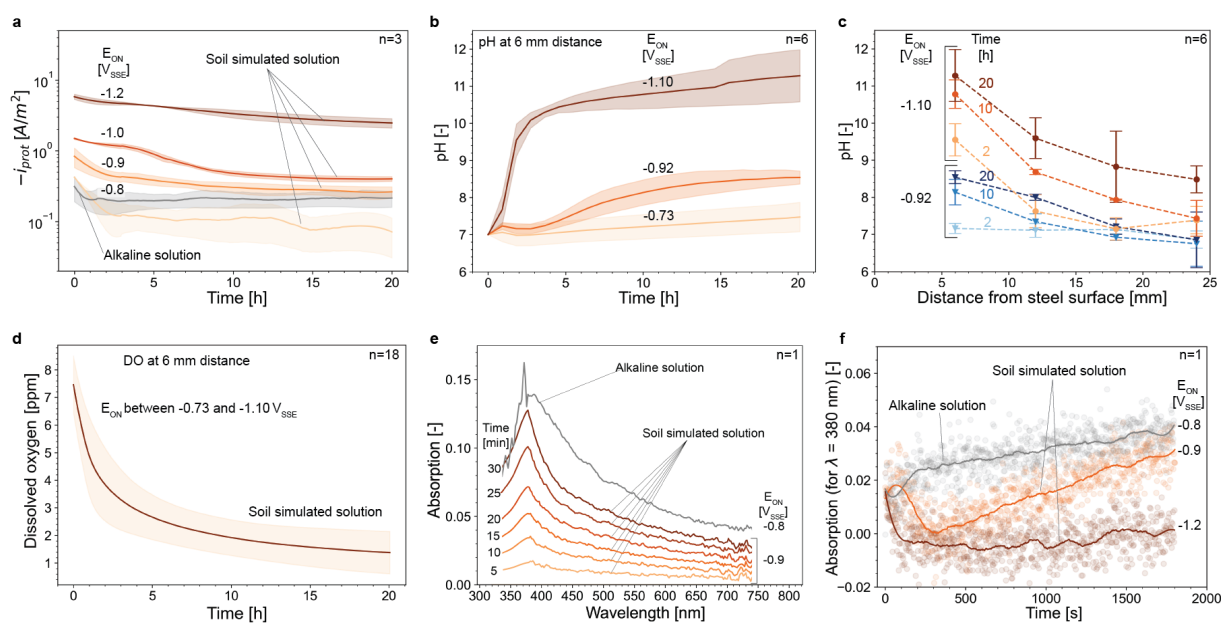

*Figure 1: a) Mean (line) and standard deviation (shade) of the protection current densities applied for the different $E_{ON}$ and electrolytes. b) Mean (line) and standard deviation (shade) of the in-situ pH measurements at 6 mm distance from the WE surface as a function of the applied $E_{ON}$ and experimental time. c) Mean and standard deviation of the pH measurered upon CP as a function of the distance from the WE surface, experimental time, and $E_{ON}$. d) Mean (line) and standard deviation (shade) of the dissolved oxygen (DO) concentration as a function of the experimental time for $E_{ON}$ between -0.73 and -1.1 $V_{SSE}$ measured at 6 mm from the surface of the WE. e) Light absorption spectra over time for the samples polarised to -0.8 $V_{SSE}$ and -0.9 $V_{SSE}$ in alkaline and soil simulated solution, respectively. f) Absorption of the light with wavelength 380 nm for samples under polarisation in alkaline and soil simulated solution. The number of samples (n) used for the statistical analysis is indicated in each figure.*

*Ex-situ surface characterisation*

While in-situ methods allow us to identify the presence or absence of a supposed iron oxide film and monitor its growth over time, ex-situ methods allow obtaining quantitative information and mapping the chemical composition of the interface of interest upon termination of the experiments. FIB-TEM, STEM and EDS was conducted on carbon steel, ex-situ, after 20 hours of polarisation in either simulated soil solution or alkaline solution at different $E_{ON}$. The efficacy of our sample preparation protocol was tested by viewing a freshly polished steel sample under the TEM (Supplementary Figure 1) and was validated by the absence of an oxide film at the interface between the freshly polished steel surface and the Pd/Pt coating. The TEM/STEM/EDS micrographs (Figure 2 and Supplementary Figure 1) show the presence of an oxide film (rich in Fe and O from the EDS maps) on the steel surface after polarisation to potentials -0.8 $V_{SSE}$, -0.9 $V_{SSE}$, and -1.0 $V_{SSE}$ in simulated soil solution, and -0.8 $V_{SSE}$ in the highly alkaline electrolyte. For all these cases, an oxide film with a thickness of ~2-5 nm was observed. As would be expected for the case of polarising the steel at -1.2 $V_{SSE}$, in the immunity region of the Pourbaix digram[46] as mentioned above, an oxide film could not be observed. Note that this is also in line with the almost constant zero absorption at λ = 380 nm over time in the PRM measurements (Figure 1f). These ex-situ micrographs confirm that the peak in the absorption spectra at λ = 380 nm (observations from in-situ UV-Vis spectroscopy (Figure 1e)) is related to the presence of an iron oxide film.

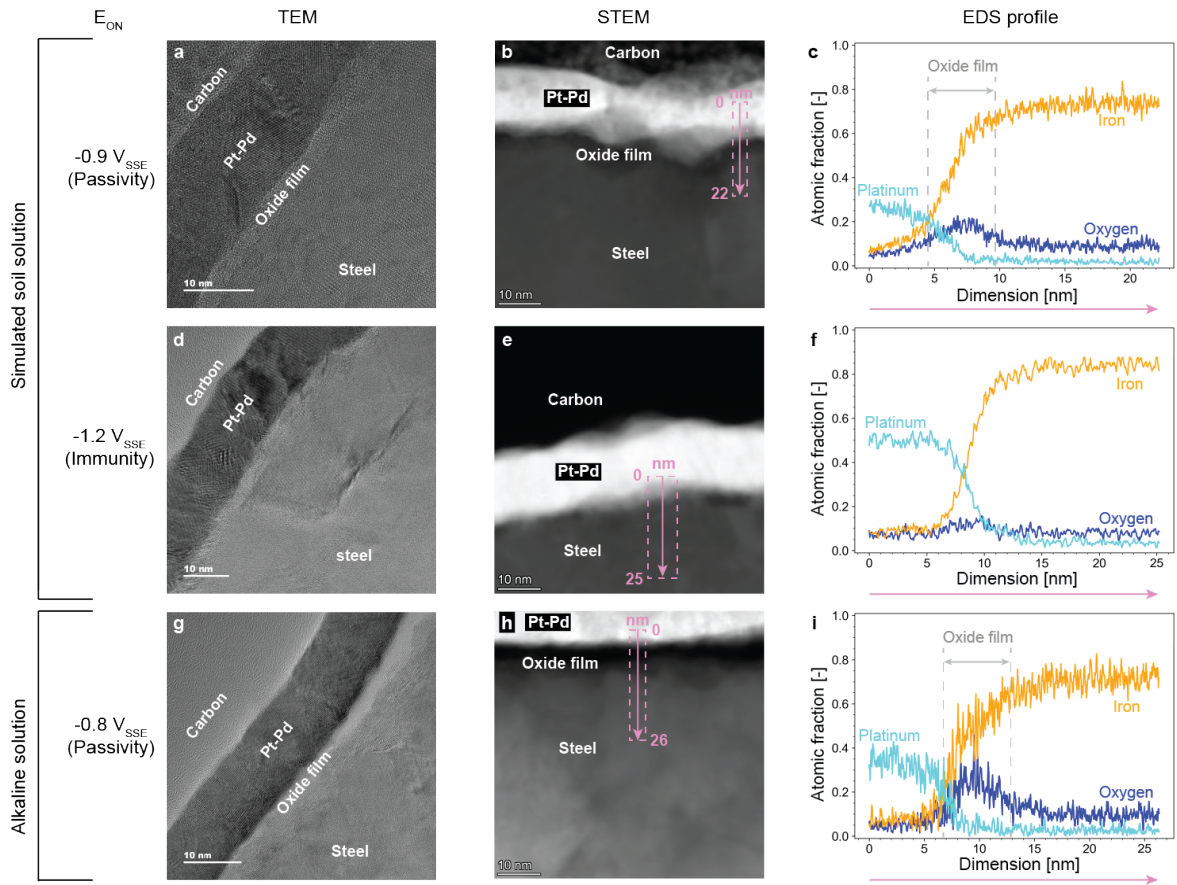

*Figure 2: Ex-situ TEM micrographs (a,d,g), high-angle annular dark-field STEM micrographs (b,e,h) and EDS profiles of iron, oxygen, and platinum (c,f,i). The location and direction of the EDS profiles is indicated on the STEM-EDS maps by means of purple arrows; the dashed box indicates the averaged width. The data was obtained for samples polarised for 20 hours to $E_{ON} = -0.9$ $V_{SSE}$ (a,b,c) and $E_{ON} = -1.2$ $V_{SSE}$ (d,e,f) in simulated soil solution as well as $E_{ON} = -0.8$ $V_{SSE}$ in alkaline solution (g,h,i).*

*Electrochemical impedance spectroscopy*

We assessed the electrochemical properties of the metal-solution interface over time, when immersed in simulated soil solution and maintained at different $E_{ON}$, using electrochemical impedance spectroscopy (EIS). In general, the EIS measurements varied significantly depending on the $E_{ON}$ and over experimentation time. For all $E_{ON}$, an increase in the real and imaginary parts of impedance can be observed from the Nyquist plots (Figure 3) over time. From the Lissajous plots aquired at 0.01 Hz (Figure 3a) which is a frequency associated with negligible capacitive effects (see Supplementary Figure 2), we could observe the current density to shift to lower values over time. This is in agreement with the $i_{prot}$ in Figure 1a.

At $E_{ON}$ = -0.8 $V_{SSE}$, we found that the currents during the sinusoidal applied voltage signal included both cathodic and anodic excursions (Figure 3a), suggesting that the system was close to the open circuit potential (OCP) over the entire experimental time. This can explained by the known rapid deoxygenation of the metal surface upon application of a cathodic protection current[23,24] (Figure 1d), which is also supported by the observed fast decrease in $i_{prot}$ below levels of 0.1 A/m$^2$ (Figure 1a). Given that this system was close to the OCP, we estimated the polarisation resistance (Supplementary Note 1) from the slope of the Lissajous plots (acquired at 0.01 Hz – Figure 3a). Conversion to corrosion current density suggests the latter to be of the order of ~70 μm/y (Supplementary Note 1), which is clearly higher than protection criteria in different standards[40,41]. It can therefore be inferred that at $E_{ON}$ = -0.8 $V_{SSE}$ cathodic protection was not fully achieved and the metal corroded at non-negligble values, however at a much lower rate than a non polarised steel exposed to a similar electrolyte[42]. We observed remarkably different electrochemical impedance spectra for the cases at $E_{ON}$ = -0.9 $V_{SSE}$ and $E_{ON}$ = -1.0 $V_{SSE}$ and these show a significant increase in impedance over time, with real impedances reaching higher than 4000 Ω cm$^2$ after 20 h. At $E_{ON}$ = -0.9 $V_{SSE}$, there seems to be a second time constant at low frequencies appearing with time, although the data geneally becomes relatively noisy after 5-10 h, which hampers the interpretation at frequencies lower than 0.1 Hz. For the case at $E_{ON}$ = -1.2 $V_{SSE}$, the Nyquist plots (Figure 3e) showed perfect semicircles, especially after 10 h of plarisation, and while the impedance increased over time, it was much lower (800 Ω cm$^2$ at 20 h) compared to all other studied polarisation conditions, suggestive of an ideal capacitive metal/solution interface. In line with the observations from PRM (Figure 1f) and FIB-TEM (Figure 2), at $E_{ON}$ = -1.2 $V_{SSE}$, the carbon steel was virtually free of an oxide film. For all other polarisation conditons ($E_{ON}$ = -0.8 $V_{SSE}$, $E_{ON}$ = -0.9 $V_{SSE}$ and $E_{ON}$ = -1.0 $V_{SSE}$), the semicircles were depressed (Figure 3b-d) and indicative of a corroding or passive iron system[43], which again agrees with the in-situ and ex-situ surface characterisation results presented above.

Conducting EIS in conjunction with polarising the carbon steel at significantly cathodic potentials ($E_{ON}$ = -1.0 $V_{SSE}$ and -1.2 $V_{SSE}$) allowed us to extract interesting information from the Lissajous plots at frequency 0.01 Hz, where capacitive effects can here be considered negligible. At these cathodic potentials, the potential-current response of the system can be assumed to be governed by the kinetics of the cathodic reaction and the Lissajous plots can be considered to represent a small portion of the cathodic reaction branch. For instance, the slope of the Lissajous plots can be used as a proxy of the cathodic Tafel slope ($b_c$, Supplementary Note 1). Given that the concentration of dissolved oxygen at the steel surface decreases rapidly to low values (Figure 1d), it is reasonable to assume that this cathodic Tafel slope over time corresponds increasingly to the cathodic reaction kinetics of the HER. At $E_{ON}$ = -1.0 $V_{SSE}$, $|b_c|$ increased with time, approaching a value of 0.45 V/dec (Supplementary Note 1), which is in line with litetature values for HER on carbon steel with an iron oxide film[44]. The change in $b_c$ over time suggests increasingly limited HER kinetics, which agrees with our results on time-dependent $i_{prot}$, pH, dissolved oxygen, PRM (Figure 1) and the FIB-TEM (Figure 2) confirming the presence of an iron oxide film. On the other hand, at $E_{ON}$ = -1.2 $V_{SSE}$, the opposite trend for $|b_c|$ was observed, namely a decrease in $|b_c|$ towards values around 0.3 V/dec (Supplementary Note 1). This may be explained by the fact that the $|b_c|$ of HER is lower on bare iron compared to iron covered with an oxide film[44], and that at $E_{ON}$ = -1.2 $V_{SSE}$, the steel electrode is in the immunity domain and is free from an oxide film (Figure 2).

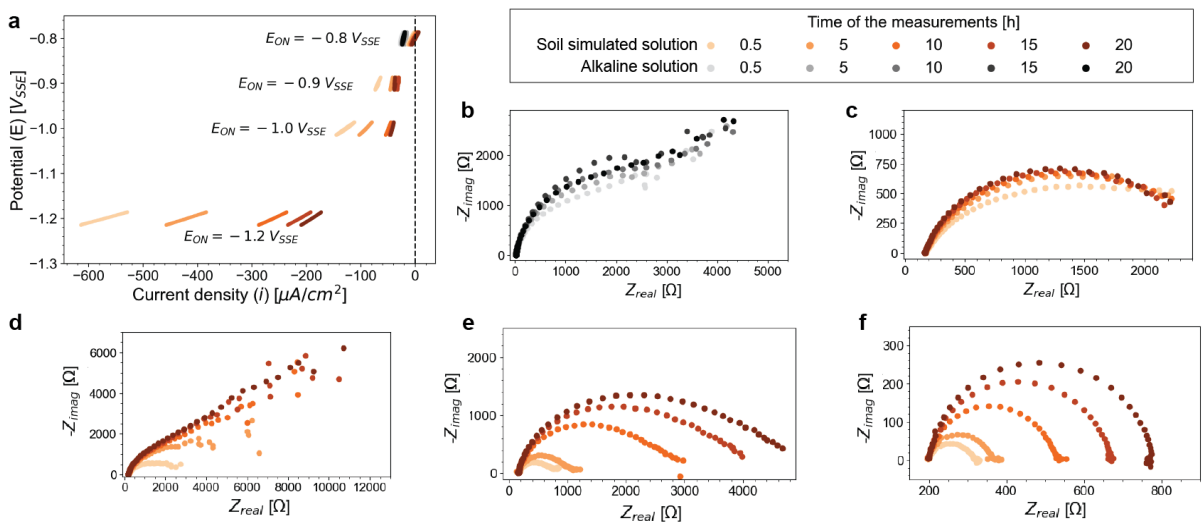

*Figure 3: In-situ EIS measurements of carbon steel samples in simulated soil solution and an alkaline solution, a) Lissajous plots acquired for 0.01 Hz at different $E_{ON}$ as a function of the experimental time. Nyquist plots measured on carbon steel under cathodic polarisation as a function of the experimental time for b) $E_{ON}$ = -0.8 $V_{SSE}$ in alkaline solution, and c) $E_{ON}$ = -0.8 $V_{SSE}$, d) $E_{ON}$ = -0.9 $V_{SSE}$, e) $E_{ON}$ = -1.0 $V_{SSE}$, f) $E_{ON}$ = -1.2 $V_{SSE}$ in soil simulated solution.*

# Discussion

Based on our investigations of both spatio-temporal changes in the electrolyte and of the steel surface, we provide a closure to the long-standing debate on the working principle of CP. A century after Davy's experiments to protect copper in seawater from corrosion, CP of iron and mild steel in soil – a porous medium – was studied by Bauer and Vogel in Germany in the 1910s[45] and extensively by Kuhn in the US in the 1920s[46]. From the 1950s onwards, CP was also applied to protect reinforcing steel corrosion in another porous medium, namely concrete[47]. CP of iron-based alloys exhibits fundamental differences from CP of copper, amongst other reasons, because iron in the water system cannot be as easily polarised to the immunity domain as in the case of copper[39]. While the work by Bauer and Vogel[45] was largely inspired by Davy's approach and focused on varying the area ratio between sacrificial anodes such as zinc and the protected iron, Kuhn[46] carried out more rigorous experiments, striving to elucidate the underlying mechanism. Already in 1928, Kuhn hypothesised – notably without any direct experimental evidence – on the role of changes in the interfacial electrolyte chemistry by stating that the cathodic current "causes a film of hydroxide to form which protects these areas from corrosion" (in agreement with Eq. 1 and 2, shown below). Later on, scholars[9,21–25,29,30,38,48–51] have attempted to experimentally quantify this modification of the electrolyte upon CP, which was later interpreted in the context of pH-dependent stability diagrams that were proposed by Pourbaix in the middle of the last century[39]. This combination of experimental evidence for interfacial pH increases upon CP with thermodynamic stability diagrams, made a number of scholars hypothesise about the formation of an iron oxide film that then protects the steel from corrosion[11,25,28,31,37,38,52–58]. Note that in the literature, and especially in the work of Pourbaix, such films forming within the potential-pH domain where iron oxides are thermodynamically stable, are termed "passive film", because the film significantly modifies the anodic reaction kinetics, namely slowing down the iron dissolution rate by several orders of magnitude, thus promoting corrosion protection. However, direct experimental evidence for such oxide films was lacking until today. Another camp of scholars remained critical about a oxide film forming under cathodic polarisation and instead supports the view that the mechanism of CP of steel occurs through the negative shift in potential along the Tafel line of the activation-controlled iron dissolution reaction (Fe → $Fe^{2+}$ + 2$e^-$), thus reducing the rate of anodic iron dissolution[13,17,19].

We conclude that both these theories need be considered complementary to each other, rather than contradictarily, to fully elucidate the working mechanism of CP as highlighted in Figure 4a-c and discussed in the following. Prior to application of a cathodic current, at t < 0 (Figure 4a), the anodic dissolution of Fe is sustained by the oxygen reduction reaction (ORR, Eq. 1), generally under diffusion limited conditions. The released aqueous Fe hydrolyses and precipitates as Fe(II) oxides or, upon oxidation, as Fe(III) oxyhydroxides. As shown in Figure 4b, in the early stage of the application of a cathodic current, at t ≥ 0, our results indicate that the pH of the electrolyte in the vicinity of the steel increases (Figure 1b,c) and that this local alkalinsation of the electrolyte is accompanied by the removal of dissolved oxygen (Figure 1d). Eq. 1 shows the simultaneous consumption of oxygen and generation of OH-, occurring for ORR, and leading to a decrease of oxygen concentration and an increase of pH at the metal surface, which over time, due to diffusional transport, leads to concentration profiles extending into the porous medium as schematically illustrated in Figure 4b and apparent from the experimental results (Figure 1b,c). With time and depleting dissolved oxygen (Figure 1d), the ORR is overtaken by the hydrogen evolution reaction (HER) as the cathodic reaction, further resulting in the generation of OH- (Eq. 2).

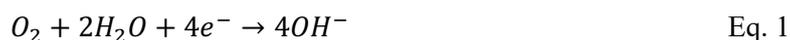

$$O_2 + 2H_2O + 4e^- \rightarrow 4OH^- \qquad \text{Eq. 1}$$

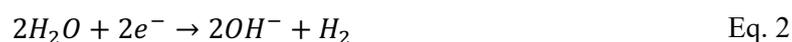

$$2H_2O + 2e^- \rightarrow 2OH^- + H_2 \qquad \text{Eq. 2}$$

A rise in the local pH (Figure 1b,c), modifies the thermodynamic stability of the polarised steel-electrolyte interface (as viewed through the Pourbaix diagram for Fe-$H_2O$ – Supplementary Figure 5) from the active domain to the passive domain, facilitating the growth of an iron oxide film with composition predominantly resembling that of magnetite ($Fe_3O_4$)[39] (Figure 4c). The formation of such an oxide film has been experimentally shown both with in-situ (Figure 1e,f) and ex-situ (Figure 2,

Supplementary Figure 1) measurements, and quantified at a thickness of ~2-5 nm. Our time-resolved in-situ measurements (Figure 1e,f) suggest that the oxide film grows as the pH at the metal-electrolyte interface increases over time (Figure 1b,c). The fact that the measured absorption increases (Figure 1f) significantly faster for steel when cathodically polarised in an initially (t=0) already alkaline solution than in the case when the steel is cathodically polarised in an initially near-neutral solution (and consequentially leading to an alkaline solution at the steel-electrolyte interface), further corroborates our hypothesis that the formation of an iron oxide film is a direct consequence of the increase in pH.

The application of cathodic polarisation not only changes the electrolyte chemistry and the surface state of the steel, but also simultaneously influences the corrosion kinetics, as indicated by the marked decrease in the $i_{prot}$ observed over time (Figure 1a). We explain this decrease in $i_{prot}$ with a schematic Evans diagram – as illustrated in Figure 4d,e. Let us first understand how CP influences the cathodic reaction kinetics. Upon the application of a cathodic current (t ≥ 0), the electrolyte at the steel-electrolyte interface becomes alkaline and depleted in dissolved oxygen within a few hours (Figure 1), and, as a consequence as discussed above, the dominating cathodic reaction changes from ORR to HER. In systems with low bulk oxygen concentration, e.g. in wet soil, this transition will obviously occur faster. In any case, the local alkalinsation of the electrolyte due to ORR and HER at the steel-electrolyte interface modifies the reversible potential of hydrogen ($E_{H_2}^{rev}$) to more negative values[39]. The cathodic Tafel slopes obtained from the Lissajous plots, and especially their evolution over time (Figure 3a and Supplementary Note 1), support the suggested changes in the cathodic reaction, from ORR to HER, and the growth of an oxide film on the metal surface with time (Figure 4c,d).

Now, we consider the changes in the anodic reaction kinetics upon cathodic polarisation. The oxide film formed due to the alkaline pH (> 9) at the steel-electrolyte leads to a drastic change in the anodic reaction kinetics, as evidenced by the conversion of the anodic reaction from iron dissolution (Fe → $Fe^{2+}$ + 2e⁻) to magnetite formation (3Fe + $4H_2O$ → $Fe_3O_4$ + 8e⁻ + $8H^+$)[59]. With this transition, the reversible potential of the anodic reaction ($E_{Fe}^{rev}$) becomes pH dependent[39], and the Tafel slope increases significantly, reflecting the hindered anodic dissolution kinetics. By considering these dynamic changes in the anodic and cathodic reaction kinetics, the composition (particularly pH) of the electrolyte, the concentration of dissolved oxygen at the steel-electrolyte interface and the surface state of the steel, we can explain the decrease in the $i_{prot}$ over time (Figure 4d). Our results suggest stabilisation at longer durations between 0.05 A/m² and 4 A/m² (depending on the cathodic potential applied).

However, as one would expect, the growth of the iron oxide film would only be possible when the pH is sufficientl high and the potential is above $E_{Fe}^{rev}$, for there to be a residual anodic current density (Figure 4d,e), and not to force the steel-electrolyte interface into the immunity region of the Fe-$H_2O$ Pourbaix diagram (Supplementary Figure 5). Finally, it should be noted that if one assumed that the anodic and cathodic kinetics are constant (that is, not affected by the changes due to the cathodic polarisation, and remaining as represented for t=0 in Figure 4d also for t>0), the experimental observations of a decrease in *i*<sub>prot</sub> over polarisation time could not be explained.

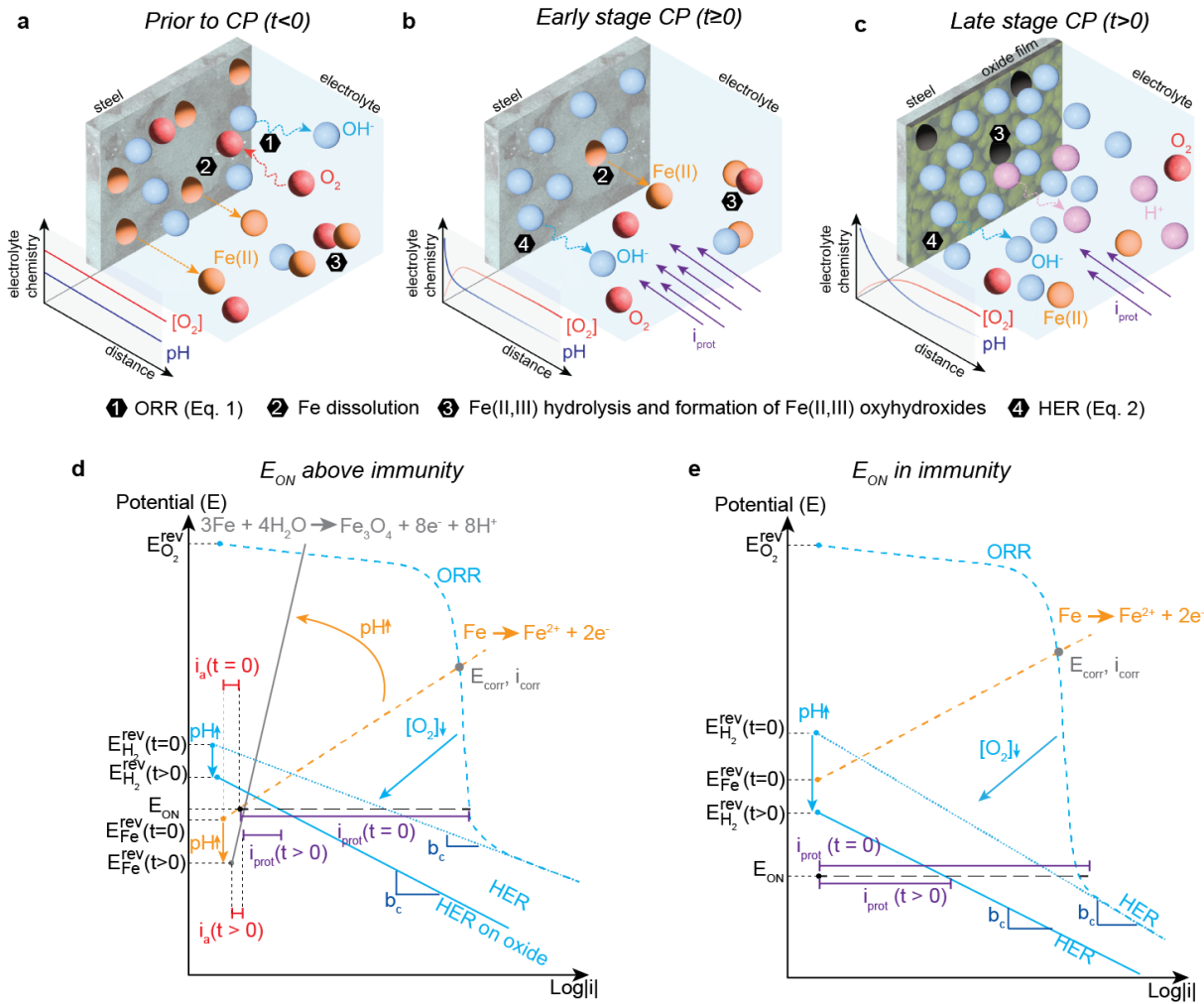

*Figure 4: Schematic representation of the cathodic protection mechanism for iron and carbon steel before the application of CP (a), in the first instants of CP application (b), and later during CP application (c). Schematic representation of the Evans diagram for iron or carbon steel in passive domain (d) and in immunity (e) in simulated soil solution with an initial pH 7.3. The kinetics reported here refer to the Pourbaix diagram. The blue lines correspond to the cathodic reactions (Eq. 1 and Eq.2) and the orange lines correspond to the anodic reaction (iron dissolution). Dotted lines represent time (t) equal to 0, namely the moment of the application of the cathodic current, continuous lines are at t>0, that is, a time where the application of the cathodic current has led to an increase in pH and a decrease in oxygen concentration at the metal surface. The dashed black lines correspond to $E_{ON}$, the purple line correspond to the applied cathodic current density ($i_{prot}$), while the red line corresponds to the residual anodic current density ($i_a$). $E_{corr}$ and $i_{corr}$ correspond to the open circuit potential (OCP) and corrosion current density, respectively, in absence of polarisation. $b_c$ represents the cathodic Tafel slope on bare metal and on iron oxide.*

This study has characterised a number of simultaneous and intertwined processes occuring at the steel-porous medium interface upon the application of CP, namely, alkalinisation and deoxygenation of the interfacial electrolyte and the formation of an iron oxide film on the steel surface. Whilst our results coherently underpin the proposed mechanism of CP of steel in porous media summarised in Figure 4, it is important to note that the thermodynamic stability domains of immunity, active corrosion, and passivity depends on the activity of Fe and the presence of species such as chloride, sulfates, or carbonates. Thus, the here observed relationships between applied potential ($E_{ON}$), resulting protection current, pH, oxygen concentration, and growth of oxide film, are not directly applicable to other conditions. In particular the effect of complexing species such as carbonate ions may have an important influence on $E_{ON}$ needed to ensure corrosion protection (compare Supplementary Note 2)[60].

The findings and mechanistic insight of this study contribute to enhancing corrosion protection technologies and resolving some of the severe limitations of the state-of-the-art engineering practices related to corrosion in safety-relevant steel-based structures, such as in the energy and construction sector. Criteria for the assessment of the effectiveness and to ensure the safe operation of CP have been debated for almost a century[11,12,18,29,49,61–63], with the controversy arising primarily from a lack of scientific comprehension of the mechanism of CP. It should also be mentioned that protection criteria in standards are largely based on empiricism rather than on scientifically rigorous concepts[10,11,49,64]. Interestingly, our findings now support the pausibiliy and correctness of certain concepts in these standards (Supplementary Note 3). Nevertheless, a severe shortcoming of state-of-the-art engineering is that the requirements specified in various standards sometimes conflict with one another, making it difficult to satisfy all relevant standards simultaneously[10,64]. To overcome these limitations of engineering practice, new approaches have been proposed[64]. It is important to recognise that these approaches are based on the previously hypothesised concept that an iron oxide film forms upon CP [10–12,16,20,26]. In this context, our findings serve as a crucial basis for future engineering practice to devise consistent standards that are not in conflict with each other and are based on parameters that can be measured on modern and well-coated structures.

# Materials and Methods

## 1.1 Materials and setups

Carbon steel (grade 1.0330, with C < 0.12%, Mn < 0.6%, P < 0.045% and S < 0.045%, according to the specifications of the supplier) was used as the working electrode (WE) in this study. The reference electrode (RE) was Ag/AgCl sat. KCl (SSE) and the counter electrode (CE) was titanium metal mixed oxides (TiMMO) mesh. NaOH pellets (≥ 98 %), anhydrous $Na_2SO_4$ (> 99 %), $NaHCO_3$ (> 99.7 %) and Ethylenediaminetetraacetic acid disodium salt dihydrate (EDTA) (> 99 %) of ACS reagent grade were purchased from Sigma Aldrich. For all the experimental setups used in this study, stirring of the electrolyte was not implemented.

*Setup for steel characterisation*

The WE was a cylindrical carbon steel bar with its lateral surface coated with epoxy resin, which was applied with fluidised bed epoxy coating technique. With this process, possible gaps between the steel and the epoxy resin are minimised, thereby, reducing the risk of crevice corrosion. After soldering the back side of the sample to a copper cable to provide electrical connection, the sample was embedded in an acrylic resin, protecting the cable connection at the back side as well as the lateral surface. The uncoated front surface of the bar, which has a cross-sectional area of 1 cm$^2$, was polished down to 1 μm with a diamond paste and then exposed to electrolytes for all subsequent tests. Between polishing and immersion in the electrolyte, the sample was stored for 30 mins in a desiccator containing silica gel (i.e. relative humidity about 10%).

Two electrolytes were used in this study: an alkaline solution composed of 0.1 M NaOH + 2 mM EDTA with pH equal to 13, and a simulated soil solution composed of 5 mM $Na_2SO_4$ + 2.5 mM $NaHCO_3$ + 2 mM EDTA with pH equal to 7.3. EDTA is well known to have a strong affinity for aqueous iron[65,66] and reduce the availability of aqueous iron to participate in other chemical reactions[65–67]. EDTA was added in the electrolytes to minimise the precipitation of iron hydroxides at the metal surface[68,69], which could potentially interfere with in-situ surface characterisation. EDTA is generally not present in natural environments, but other species (i.e., carbonate ions, $CO_3^{2-}$ [48,60]) with similar affinity to iron might be present in natural soils.

The pH of the solutions was measured by means of Mettler-Toledo SevenExcellence pH meter in conjunction with a glass membrane electrode. The pH meter was calibrated against commercial (Mettler-Toledo and Sigma-Aldrich) buffer solutions of pH 7, 11 and 13. The measurements in alkaline solutions were performed relatively quickly to reduce the error as a consequence of the release of silica from the glass membrane.

*Setup for electrolyte characterisation*

In order to limit the setup to a 1D problem, an electrochemical cell with internal dimensions 20x20x200 mm$^3$ was designed to study the temporal and spatial modifications in the soil chemistry during the application of CP (Supplementary Figure 4).

The dimensions of the WE for this setup were 19x19x2.5 mm$^3$. The back side of the WE was soldered to a copper wire to provide electrical connection. Subsequently the WE was embedded in acrylic resin to isolate the back side as well as the lateral side from the electrolyte. The WE was polished down to 5 μm with SiC grinding paper leaving an exposed area of 3.61 cm$^2$. The WE was then located on one of the sides of the electrochemical cell. On the opposite face of the cell, the CE was positioned at a distance of ~195 mm from the surface of the WE. The RE was positioned 6 mm away from the surface of the WE to limit the contribution of IR-drop during the measurements.

Potentiometric sensors (Duramon AG), to measure pH in-situ, were positioned at 6 mm, 12 mm, 18 mm, and 24 mm from the WE surface. Additionally, an optical sensor (Pyroscience GmbH) was placed at 6 mm from the WE surface to monitor dissolved oxygen (DO) concentration. The location of the pH and oxygen sensors allowed monitoring temporal and spatial changes in the chemistry of the electrolyte in proximity of the WE upon the application of CP.

For these experiments the electrolyte was composed by quartz sand with granulometry lower than 0.9 mm saturated with a solution composed by 5 mM $Na_2SO_4$ + 2.5 mM $NaHCO_3$ and adjusted to a pH of 7 by adding HCl.

Before the experiment, the quartz sand was carefully washed with dionised water and dryed in oven at 80°C multiple times until the conductivity of the washed out water was in the same range as the conductivity of the dionised water. These iterations were performed to remove possible contaminants that could comprimise the chemical composition of the solution.

## 1.2 Electrochemical polarisation and measurements

The electrochemical conditioning and measurements were performed with Gamry interface 1010E. On all samples, cathodic polarisation was applied in potentiostatic mode up to 20 hours. Different ON-potentials ($E_{ON}$) in the range of -0.73 $V_{SSE}$ and -1.20 $V_{SSE}$ were applied. The corresponding cathodic protection current densities ($i_{prot}$), were recorded over time. The electrochemical cell used for these electrochemical polarisation tests and measurements combined with PRM and UV-vis spectroscopy is described elsewhere[48,70].

Additionally, companion samples were identically polarised in a conventional three-electrode electrochemical cell, which allowed for in-situ characterisation by means of electrochemical impedance spectroscopy (EIS). EIS measurements were repeated once with a fresh sample for each configuration. As the results showed very good reproducibility, data pertaining to only one sample is presented in this paper.

Upon polarising the sample cathodically, EIS was conducted every 30 mins for 20 hours. To this aim, a sinusoidal voltage signal was superimposed on different $E_{ON}$. The superimposed root-mean-squared voltage signal ($V_{RMS}$) was 15mV. The frequency range used to study EIS was between 5000 Hz and 0.01 Hz.

## 1.3 Steel and electrolyte characterisation techniques

*UV-vis (in-situ)*

UV-vis measurements were carried out using a Lambda 650 UV–Vis spectrophotometer from Perkin Elmer. The absorption spectra were recorded between wavelengths 800 and 320 nm, with step size of 2 nm. The electrochemical cell was located at the reflection station of the UV-vis spectrophotometer, and the WE was positioned in correspondence to the center of the incident light window (with a diameter about 2.5 cm). The WE surface was positioned 4 mm from the Suprasil glass of the electrochemical cell, making the effective path length of light roughly 8 mm. The reflected and scattered light were collected by the integrating sphere and directed to the detector.

*Photometric reflectance measurements (in-situ)*

For photometric reflectance measurements (PRM), a Deuterium lamp (30 W) was used as the light source, providing a continuous light with wavelength between 200 nm and 700 nm (UV and visible range). An optical fiber with diameter of 400 μm and 2 m in length guided the light towards the WE surface. Ferrule clamps and a 3D printed support in polylactic acid (PLA) allowed positioning the optical fibers at a distance of 6 ± 0.15 mm from the WE, with an incident ($α_i$) and reflection angle ($α_r$) of 30° ± 0.2°. The diameter of the light beam on the WE surface was about 2 mm. The reflected light was collected by means of a second optical fiber and guided to the spectrometer (Thorlabs CCS200). The spectrometer and the potentiostat were connected via USB to a computer and controlled with the dedicated software. Detailed description of the setup has been reported elsewhere[48,70].

The absorption of the film was calculated with Eq. 3[69]:

$$Absorption = 1 - \frac{I(\lambda)}{I_0(\lambda)} \qquad \text{Eq. 3}$$

where, *I* is the intensity of the reflected light, and $I_0$ is the intensity of the reflected light on the oxide-free (initial) surface. The integration time of 2.5 s was selected in order to maximise the signal to noise ratio in the UV region. As observed from UV-vis measurements, the most significant modification in absorption was measured at 380 nm, and therefore we chose to evaluate the PRM data at that wavelength.

*FIB-TEM (ex-situ)*

Samples that had been polarised for 20 hours in the test solutions under different conditions were retrieved, immediately dried with $N_2$ (g) and stored in a desiccator containing silica gel crystals under a $N_2$ atmosphere. For comparison, a freshly polished sample underwent the same protocol. Subsequently, the samples were coated with Pt-Pd with a thickness of 5 nm within 2 hours from the extraction from the testing solution in order to limit any further oxidation. Then, TEM lamellae with dimensions about $12 \times 5 \times 0.08$ μm$^3$ were extracted from the WE.

The lamellae were prepared by focussed ion beam (FIB) (Helios 5 UX, Thermo Scientific, the Netherlands) using AutoTEM 5 software (Thermo Scientific, the Netherlands). A schematic representation of the sample preparation for TEM analysis can be found in Supplementary Figure 3. A protective carbon layer was deposited on the region of interest (ROI) first by an electron beam (2 kV, 13 nA), followed by an ion beam (30 kV, 1.2 nA). The chuck milling and lamellae thinning were done at 30 kV with FIB current from 9 nA to 90 pA with gallium ions. The lamellae were then polished at 5 kV (17 pA) and finished at 2 kV (12 pA). TEM investigations were performed at Talos F200X (FEI, the Netherlands) operating at 200 kV. Both TEM and scanning transmission electron microscope (STEM) modes were used. Energy dispersive spectroscopy (EDS) mapping was acquired by using a Super-X EDS system in the STEM mode.

*pH measurements (in-situ)*

IrOx sensors (supplied by Duramon AG) were used to monitor the pH variations during CP conditions allowing a temporal and spatial resolution of the pH. The equilibrium potential of these IrOx sensors at a given temperature depends only on the pH of the solution. Each sensor was subjected to a 5 points calibration in buffer solutions in pH range between 7 and 13 before and after every experiment. With IrOx sensors the changes in pH could be measured at different distances from the steel surface during the application of CP.

*Dissolved oxygen concentration measurements (in-situ)*

The oxygen concentration was monitored by fiber optical oxygen sensors. The oxygen probe (OXROB3) and datalogger (Optical Oxygen Meter - FireStingO2) from Pyroscience GmbH were used to monitor the oxygen concentration in proximity to the WE surface during the application of CP. The sensors provided information on the oxygen consumption in the vicinity of the WE surface (6 mm) over time. The calibration of the oxygen sensor was performed before each experiment with a procedure suggested by the manufacturer.

69. Büchler, M. Experimental modeling of passive films on iron: investigation of semiconductive properties of passive films, synthetic iron oxides and hydroxides in combination with a new light reflectance technique. (diss. ETH No. 12504, ETH Zurich, 1998).

70. Martinelli-Orlando, F., Dénervaud, E., Grange, R. & Angst, U. Second-harmonic generation technique for in situ study of passive film formation on carbon steel surfaces in aqueous solutions. *Materials and Corrosion* **74**, (2022).

## Acknowledgements

The authors are grateful to the European Research Council (ERC) for the financial support provided under the European Union's Horizon 2020 research and innovation program (grant agreement no. 848794). Additionally, the authors acknowledge the industrial partners Gaznat SA, Gasverbund Mittelland AG, Erdgas Ostschweiz AG and Transitgas AG for their financial support. The authors thank Maximilian Ritter and Christopher Dreimol from the group of Wood Materials Science at ETH Zurich for the support with the experiments. Moreover, the authors gratefully acknowledge the Scientific Center for Optical and Electron Microscopy (ScopeM) at ETH Zurich, in particular Dr. Peng Zeng for the support with the TEM, STEM, and EDS data acquisition. Lucas Michel from the group of Durability of Engineering Materials at ETH Zurich is acknowledged for the fruitful discussions. The contributions of SM were partially made possible by the Swiss National Science Foundation (Grant: PP00P2_194812).

## Author Contributions

FM and UA conceived the experiments. FM performed all experiments and wrote the first draft of the manuscript. All authors contributed to the analysis and interpretation of the data and to the writing of the manuscript. FM created the illustrations.

## Competing Interests

The authors have no competing interest to disclose.

## Data availability

The data that support the findings of this study are available from the corresponding author upon reasonable request.

## Correspondence

Correspondence to Ueli M. Angst (uangst@ethz.ch)

# Supplementary Information

*Mechanism of cathodic protection of iron and steel in porous media*

Federico Martinelli-Orlando, Shishir Mundra, Ueli M. Angst

**Table of contents**

# Supplementary Figure 1

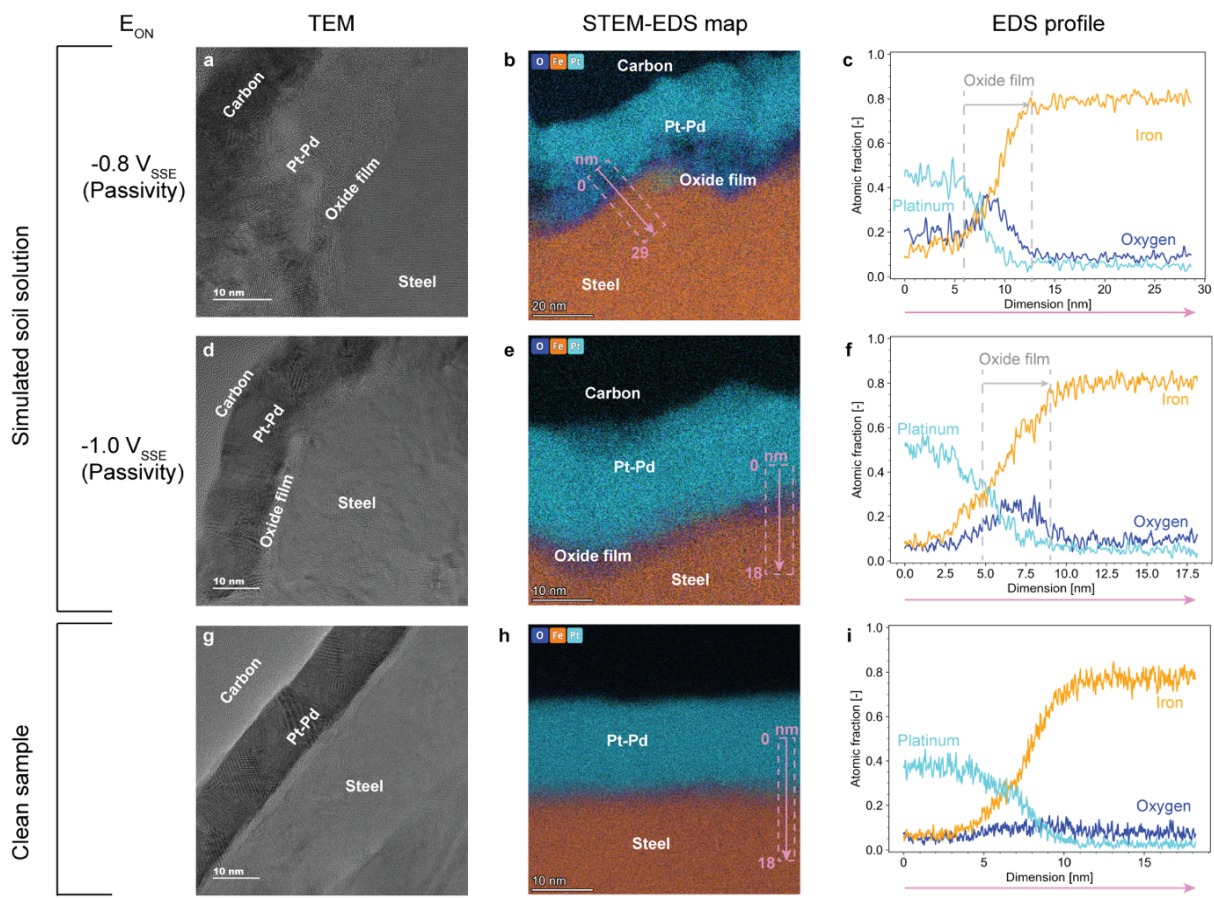

*Supplementary Figure 1: Ex-situ TEM micrographs (a,d,g), high-angle annular dark-field STEM micrographs with overlaid EDS maps of iron, oxygen, and platinum (b,e,h), and EDS profiles (c,f,i). The location and direction of the EDS profiles is indicated on the STEM-EDS maps by means of purple arrows; the dashed box indicates the averaged width. The data was obtained for samples polarised for 20 hours to $E_{ON} = -0.8\ V_{SSE}$ (a,b,c), $E_{ON} = -1.0\ V_{SSE}$ (d,e,f) in simulated soil solution, with the exception of the "clean sample" that was a freshly polished sample viewed in TEM, which was used to test our sample preparation protocol, especially to validate the absence of an oxide film potentially forming during preparation and transference of the sample to the TEM (g,h,i).*

# Supplementary Figure 2

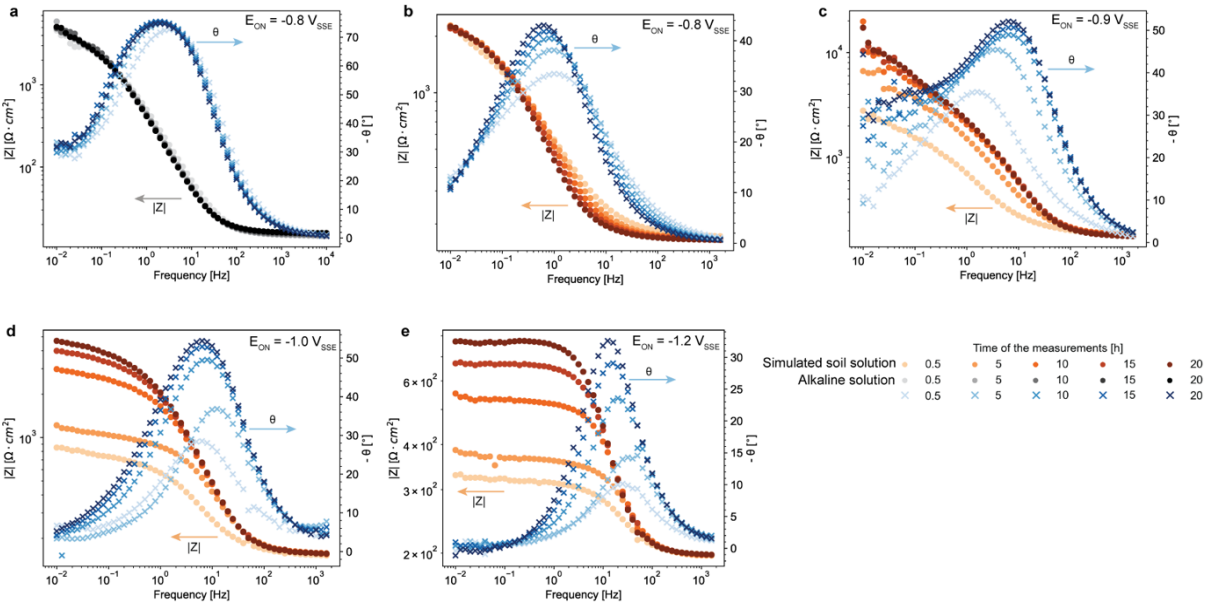

*Supplementary Figure 2: Bode plots for the sample exposed to alkaline solution (grey) and soil simulated solution (orange) for the different $E_{ON}$ applied as a function of the experimental time.*

# Supplementary Figure 3

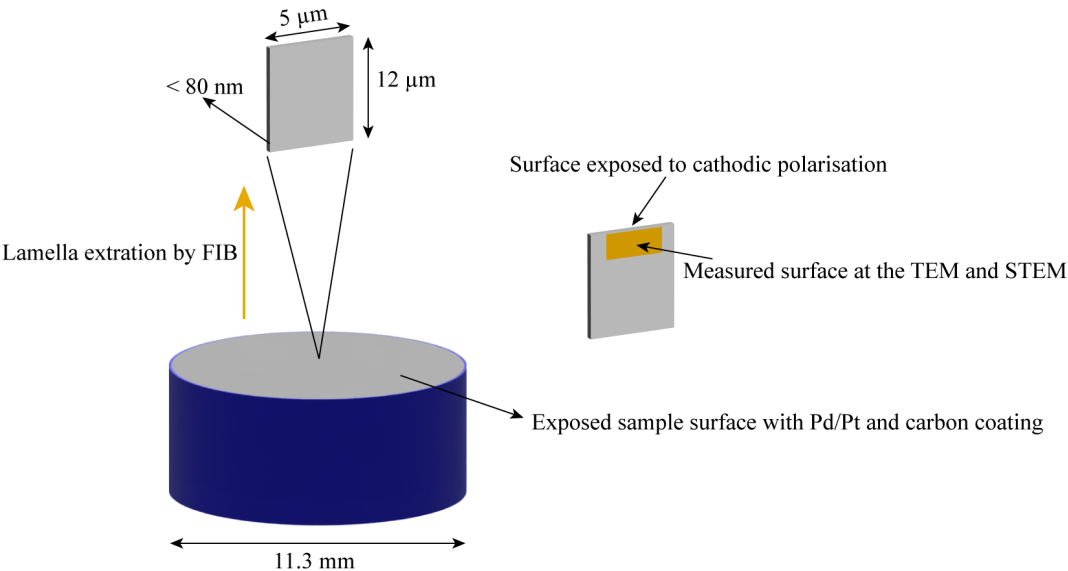

*Supplementary Figure 3: Schematic representation of the sample preparation for the analysis at the TEM and STEM*

# Supplementary Figure 4

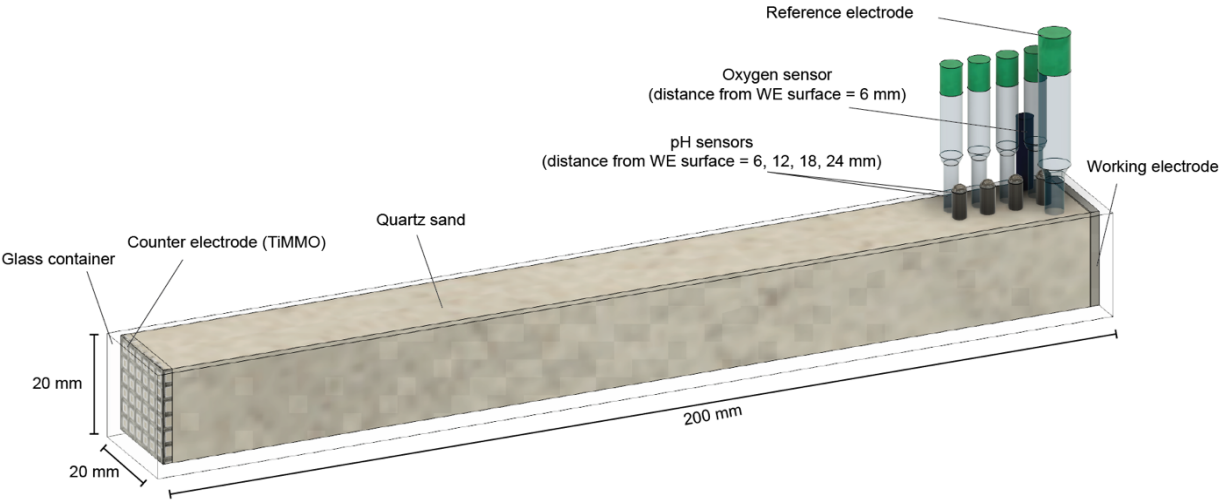

*Supplementary Figure 4: Schematic representation of the setup used to characterize the electrolyte during CP application of steel in sand.*

## Supplementary Figure 5

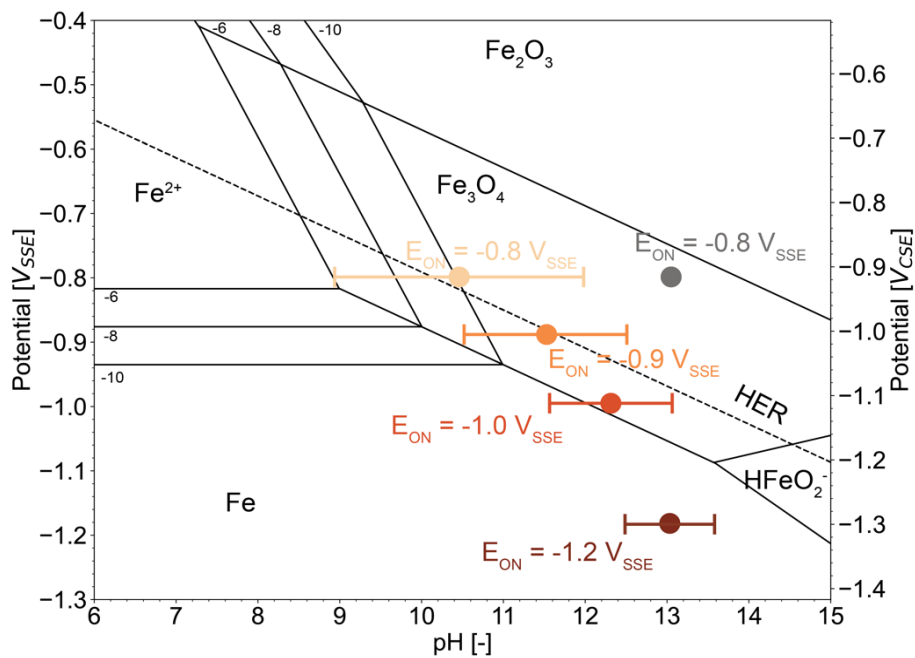

*Supplementary Figure 5: Pourbaix diagram of Fe in water at 25 °C considering a concentration of dissolved $Fe^{2+}$ in solution of $10^{-6}$ M, $10^{-8}$ M, and $10^{-10}$ M. The continuous lines represent the equilibria between the different domains, while the dashed line corresponds to the reversible potential of the hydrogen evolution reaction (HER). The dots represent the estimated location of the WE in the Pourbaix diagram, whereby the potential is known ($E_{on}$) and the pH is estimated from our experimental data and literature[1–10] (uncertainty in pH indicated by error bars). Grey dot = experiment performed in initially alkaline solution; coloured dots = experiments performed in simulated soil solution. Potentials are versus the saturated silver/silver chloride electrode (SSE).*

# Supplementary Note 1

**On the kinetics of the electrochemical reactions**

This section presents the polarization resistance ($R_p$), cathodic Tafel slope ($b_c$) and instantaneous corrosion rate, as calculated from linear fitting of the Lissajous plots taken at frequency 0.01 Hz of the electrochemical impedance spectroscopy measurements at different $E_{ON}$ and times. The corrosion rate on the basis of $R_p$ was calculated by using Stern-Geary[11,12] equation and the Stern-Geary coefficient of 13 mV [13]. The following table summarises the results.

*Supplementary Table 1: Polarization resistance (Rp), cathodic Tafel slope (b$_c$) and corrosion rate calculated from linear fitting of the Lissajous plots at different $E_{ON}$ and times.*

| Electrolyte | $E_{ON}$ [$V_{SSE}$] | Time [h] | $R_p$ [$\Omega \cdot cm^2$] | $|b_c|$ [V/decade] | Instantaneous corrosion rate [µm/y] |
|---|---|---|---|---|---|
| Simualted soil soilution | -0.8 | 0.5 | 2227 | - | 68 |
| | | 5 | 2193 | - | 69 |
| | | 10 | 2224 | - | 68 |
| | | 15 | 2162 | - | 70 |
| | | 20 | 2207 | - | 68 |
| | -0.9 | 0.5 | 2727 | - | - |
| | | 5 | 6535 | - | - |
| | | 10 | 14678 | - | - |
| | | 15 | 8670 | - | - |
| | | 20 | 9908 | - | - |
| | -1.0 | 0.5 | 846 | 0.25 | - |
| | | 5 | 1213 | 0.25 | - |
| | | 10 | 2967 | 0.34 | - |
| | | 15 | 3975 | 0.40 | - |
| | | 20 | 4658 | 0.45 | - |
| | -1.2 | 0.5 | 329 | 0.43 | - |
| | | 5 | 386 | 0.37 | - |
| | | 10 | 554 | 0.33 | - |
| | | 15 | 671 | 0.33 | - |
| | | 20 | 771 | 0.33 | - |
| Alkaline solution | -0.8 | 0.5 | 3853 | 0.22 | - |
| | | 5 | 4907 | 0.20 | - |
| | | 10 | 4298 | 0.21 | - |
| | | 15 | 4157 | 0.21 | - |
| | | 20 | 4311 | 0.22 | - |

Plotting the change of cathodic Tafel slope over time yields the diagram shown below. These results show that the cathodic Tafel slope undergoes a change. In the case of polarisation to $E_{ON}$ = –1.0 $V_{SSE}$ (blue), the Tafel slope increases with polarization time. It is our hypothesis that this change is due to i) the transition from ORR to HER as the dominating cathodic reaction, and ii) the growth of an oxide film ("passive film") on the metal surface, further modifying the kinetics of the cathodic reaction. The cathodic Tafel slope of the HER after 20 h of CP is in the range of 450 mV/dec, which is in excellent agreement with the values (416…450 mV/dec) reported by Lu et al.[14] for HER on steel covered with a passive film (range indicated in the figure with the blue shaded area). On the other hand, for polarisation to $E_{ON}$ = –1.2 $V_{SSE}$ (orange), the Tafel slope decreases with polarization time, stabilizing at values of around 300 mV/dec. Again, we suggest that these changes are the result of a transition from ORR to HER as well as the reduction of any possible scale of Fe(II, III) oxyhydroxides, leading to a virtually blank metal surface. This hypothesis is supported by literature data[15] on cathodic Tafel slopes of the HER, measured on *blank* carbon steel electrodes, being in the same range as the values found at $E_{ON}$ = –1.2 $V_{SSE}$ (Supplementary Figure 6).

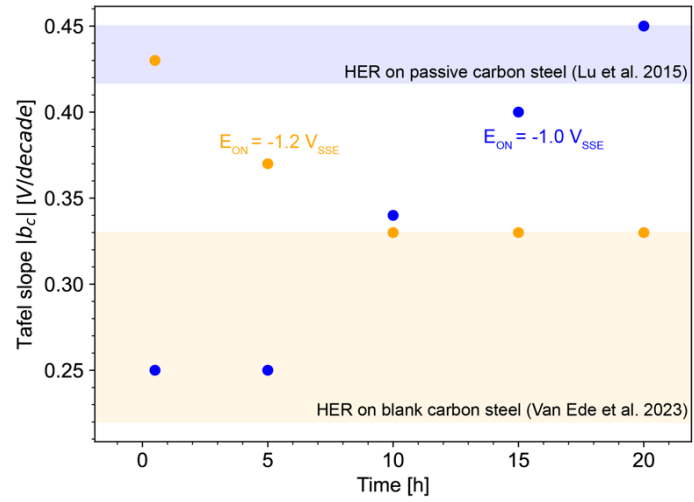

*Supplementary Figure 6: Tafel slopes of the cathodic reaction for two selected $E_{on}$ over time, as extracted from the Lissajous plots at 0.01 Hz of the electrochemical impedance spectroscopy (EIS) measurements. Shaded areas indicate literature data for the HER on passive (oxide film covered) carbon steel (blue) and blank steel (orange), see text for references.*

## Supplementary Note 2

**On the effect of complexing agents**

Ethylenediaminetetraacetic acid (EDTA) used in our experiments (at a concentration of 2 mM) is expected to complex relatively rapidly with aqueous iron, thereby decreasing the concentration of free Fe(II) below the solubility limit of Fe(OH)$_2$ [16–21], [22]. To illustrate the influence of EDTA complexing with aqueous Fe(II), we reconstructed the Pourbaix diagram (Supplementary Figure 5) for three different concentrations of free Fe$^{2+}$ (10$^{-6}$ M, 10$^{-8}$ M and 10$^{-10}$ M). In the case of the sample at $E_{ON}$ = -0.8 V$_{SSE}$ in simulated soil solution, it is likely that sample was in the active domain as long as EDTA could complex with Fe(II) (the case of 10$^{-10}$ M free Fe(II)). Once the complexation capacity of EDTA became exhausted, we would presume that Fe(OH)$_2$ might have precipitated from the solution and deposited at the sample surface (the solubility of Fe(OH)$_2$ shows a minimum between pH 10 and 12[22,23]). As a result, the boundary of the active and passive domain may have shifted to lower pH, facilitating the sample at $E_{ON}$ = -0.8 V$_{SSE}$ to be located in the passive domain, promoting the growth of an iron oxide film on the metal surface; and followed by a porous Fe(OH)$_2$ scale. It may therefore be assumed that the sample at $E_{ON}$ = -0.8 V$_{SSE}$ in simulated soil solution was at the boundary of the active and passive domains (depending on the complexation capacity of EDTA), possibly exhibiting a heterogeneous surface, with some parts active and some remaining passive. This may explain the non-negligible corrosion rates (about 50-70 µm/y – averaged over the entire surface of the sample) that were measured with EIS (Supplementary Note 1).

# Supplementary Note 3

**On the plausibility of certain concepts in engineering standards**

The findings of this study contribute to formulating a rigorous theoretical framework for the empirical understanding that had been employed to devise current industry norms, regulations and standard practices in the field of CP.

First, the concentration polarisation arising from oxygen consumption and alkalinity generation at the metal-electrolyte interface allows for explaining why **intermittent CP** (during stray current interference or when the protection current is temporarily switched off) remains protective for a considerable amount of time. That CP remains protective upon temporary stoppage of the applied current had empirically long been known, dating back even to Kuhn[24]. This, however, cannot be explained only due to changes in the kinetics of the anodic and cathodic reactions alone, but also the changes in the pH and oxygen concentration of the interfacial electrolyte in the porous medium must be considered. Consequently, our results provide the scientific basis for the "**depolarisation criteria**" employed by various standards and codes for CP of steel in soil or concrete[25–28], that rely on a potential shift occurring over hours-days.

Moreover, our approach of viewing CP as a combination of the changes in the electrolyte chemistry and the formation of an oxide film allows us to also to confirm the aptness of the present limits prescribed by SN EN ISO 21857 for **stray currents**. This is because only in the presence of a protective oxide film formed on the cathodically polarised steel, can temporarily high anodic currents be tolerated without resulting in relevant signs of corrosion.

Finally, the currently hypothesised mechanisms of **AC corrosion** on cathodically protected steel rely on the presence of an oxide film[6,29–31]. The changes induced by cathodic polarisation on the electrolyte chemistry and presence of an oxide film at the porous medium-steel interface, as observed in this study, confirms the hypothesis that a protective passive film is formed under low levels of cathodic polarisation. Furthermore anodic and cathodic excursions due to AC interference may indeed result in the dissolution and reformation of the protective passive film and lead to the generation of rust at the metal/electrolyte interface as specified in SN EN ISO 18086 as well as in DVGW GW 28 and DVGW GW 28 B1.

Thus, we provide crucial experimental proof that, for the first time, demonstrates the correctness and the plausibility of aforementioned models and standards, ensuring the integrity of safety-relevant structures, which previously relied on empirical knowledge.